\documentstyle[11pt]{article}
\def\bea{\begin{eqnarray}}
\def\eea{\end{eqnarray}}

\def\beq{\begin{equation}}
\def\eeq{\end{equation}}
\def\ba{\beq\new\begin{array}{c}}
\def\ea{\end{array}\eeq}
\def\be{\ba}
\def\ee{\ea}
\def\stackreb#1#2{\mathrel{\mathop{#2}\limits_{#1}}}

\def\2{{1\over 2}}
\def\f{1\over}
\makeatletter
\newdimen\normalarrayskip              
\newdimen\minarrayskip                 
\normalarrayskip\baselineskip
\minarrayskip\jot
\newif\ifold             \oldtrue            \def\new{\oldfalse}
\def\arraymode{\ifold\relax\else\displaystyle\fi} 
\def\eqnumphantom{\phantom{(\theequation)}}     
\def\@arrayskip{\ifold\baselineskip\z@\lineskip\z@
     \else
     \baselineskip\minarrayskip\lineskip2\minarrayskip\fi}
\def\@arrayclassz{\ifcase \@lastchclass \@acolampacol \or
\@ampacol \or \or \or \@addamp \or
   \@acolampacol \or \@firstampfalse \@acol \fi
\edef\@preamble{\@preamble
  \ifcase \@chnum
     \hfil$\relax\arraymode\@sharp$\hfil
     \or $\relax\arraymode\@sharp$\hfil
     \or \hfil$\relax\arraymode\@sharp$\fi}}
\def\@array[#1]#2{\setbox\@arstrutbox=\hbox{\vrule
     height\arraystretch \ht\strutbox
     depth\arraystretch \dp\strutbox
     width\z@}\@mkpream{#2}\edef\@preamble{\halign
\noexpand\@halignto
\bgroup \tabskip\z@ \@arstrut \@preamble \tabskip\z@ \cr}%
\let\@startpbox\@@startpbox \let\@endpbox\@@endpbox
  \if #1t\vtop \else \if#1b\vbox \else \vcenter \fi\fi
  \bgroup \let\par\relax
  \let\@sharp##\let\protect\relax
  \@arrayskip\@preamble}
\def\eqnarray{\stepcounter{equation}%
              \let\@currentlabel=\theequation
              \global\@eqnswtrue
              \global\@eqcnt\z@
              \tabskip\@centering
              \let\\=\@eqncr
              $$%
 \halign to \displaywidth\bgroup
    \eqnumphantom\@eqnsel\hskip\@centering
    $\displaystyle \tabskip\z@ {##}$%
    \global\@eqcnt\@ne \hskip 2\arraycolsep
         $\displaystyle\arraymode{##}$\hfil
    \global\@eqcnt\tw@ \hskip 2\arraycolsep
         $\displaystyle\tabskip\z@{##}$\hfil
         \tabskip\@centering
    &{##}\tabskip\z@\cr}
\begingroup\ifx\undefined\newsymbol \else\def\input#1 {\endgroup}\fi
\input amssym.def \relax
\input amssym
\newfont{\hr}{msbm10}
\newfont{\ams}{msam10}
\font\teneufm=cmmib10
\font\seveneufm=cmmib7
\font\fiveeufm=cmmib5
\def\bfit#1{{\textfont1=\teneufm\scriptfont1=\seveneufm
\scriptscriptfont1=\fiveeufm
\mathchoice{\hbox{$\displaystyle#1$}}{\hbox{$\textstyle#1$}}
{\hbox{$\scriptstyle#1$}}{\hbox{$\scriptscriptstyle#1$}}}}

\def\balpha{{\bfit\alpha}}

\def\bnu{{\bfit\nu}}

\def\blambda{{\bfit\lambda}}

\def\bn{{\bfit n}}
\def\bm{{\bfit m}}

\begin{document}
\begin{center}
\hfill FIAN/TD-12/95,\ ITEP/TH-24/96\\
\hfill hep-th/9607123\\
\vspace{0.3in}
{\Large\bf Group Theory Structures Underlying Integrable Systems
}\footnote{Talk presented at the second A.D.Sakharov International
Conference}
\end{center}
\centerline{{\large A.Mironov}\footnote{Partially
supported by grants RFFI-96-02-19085, INTAS-93-1038 and Volkswagen Stiftung}}
\centerline{{\it Theory Dept., Lebedev Physical Inst. and ITEP, Moscow,
Russia}}

\begin{quotation}
{Different group structures which underline the integrable
systems are considered. In some cases, the quantization of the integrable
system can be provided with substituting groups by their quantum
counterparts. However, some other group structures keep non-deformed in the
course of quantizing the integrable system although their treatment is
to be changed. Manifest examples of the KP/Toda hierarchy and the Liouville
theory are considered.}\end{quotation}

\bigskip

\noindent
{\bf 1.} During last decades there was a great development of integrable
theories, both classical and quantum. However, while the classical integrable
systems mostly advanced as the theory of non-linear equations, their quantum
counterparts were rather based on algebraic structures related to the
$R$-matrix. Therefore, the quantizing procedure was not always
immediate. However, the last progress in integrable theories allows one
to introduce a unified framework equally applicable to the both classical and
quantum integrable systems. This framework is based on using the group theory
structures which underline integrable system.

In fact, it was known for many years that the most elegant and effective
approach to the classical integrable systems is to use the language of group
theory [1]. However, it turns out that there are some {\it different}
group structures underlying the same integrable system. Some of the groups
act in the space of solutions to the integrable hierarchy, others can act
just on the variables of equations ("in the space-time"). In order to
quantize an integrable system, one needs just to replace the first type group
structures by their quantum counterparts. It was demonstrated [2,3]
that one can reformulate the classical non-linear
equations in these group terms, i.e. the quantizing procedure becomes really
immediate. On the other hand, the groups acting in the space-time still
remain classical even for the quantum systems.

In this short remark we would like to stress the difference between
above mentioned
group structures and to demonstrate how they can be applied. Indeed,
the groups acting in the space of solutions are used in the
course of quantizing the system. At the same time, the space-time group can
be used in the calculations of the two-point (quantum) correlation functions.
The detailed discussion of the points briefly reviewed in this note is
contained in [2,3,4].

\bigskip

\noindent
{\bf 2.} At the first part, we demonstrate how a hierarchy of non-linear
equations can be treated completely in group terms. This group acts in the
space of solutions and should be replaced by the corresponding quantum group
upon quantizing the system [2,3].

Let us start from the simplest example of the KP hierarchy. This hierarchy
is an infinite set of equations for the infinitely many functions $\{u_i\}$
of infinitely many times $\{t_k\}$. The first equation of the hierarchy (it
is the equation for the function $u_2(x,y,z)$ depending on the first three
times $t_1=x,\ t_2=y,\ t_3=z$) is the celebrated KP equation which has given
the name for the whole hierarchy:
\be
(u_{xxx}+12uu_x-4u_t)_x+3u_{yy}=0
\ee
Other equations of the hierarchy have considerably more complicated form,
more derivatives and more times involved. However, it was the striking
invention of the Japanese school (Hirota [5] and Kyoto group [6]) that
this hierarchy can be rewritten as an infinite set of equations for {\it the
only} function of infinitely many times which is called $\tau$-function.
All the equations of the hierarchy are bilinear and homogeneous. The first
one is
\be
\left(D^4_1+3D^2_2-4D_1D_3\right)\tau\cdot\tau=0
\ee
where
\be
u_2=(\log\tau)_{xx},\ \ \ \left.D_k^nf\cdot g\equiv \partial_\epsilon^k
\left\{f(t_k+\epsilon)g(t_k-\epsilon)\right\}\right|_{\epsilon=0}
\ee
Even more surprisingly, one can construct the general solution to the whole
hierarchy in terms of the free fermionic system [6]. This solution is given
by the ratio of the correlators
\be
\tau(\{t_k\})={\left<0|e^{H(t)}g|0\right>\over \left<0|g|0\right>}
,\ \ \ H(t)\equiv \sum H_kt_k
\ee
in the theory of free two-dimensional fermionic fields $\psi(z)=\sum_{{\bf
Z}}\psi_nz^ndz^{\2}$, $\psi^{\star}(z)=\sum_{{\bf
Z}}\psi^{\star}_nz^{-n-1}dz^{\2}$ with the action
$\int\psi^{\star}\bar\partial\psi$. The Hamiltonians $H_k$ giving the
infinitely many commuting flows are defined as the positive modes of the
current $J(z)\equiv \psi^{\star}(z)\psi(z)= \sum_k H_kz^{-k-1}$.

Element $g=\displaystyle{
\::\exp\left\{\sum_{m,n}A_{mn}\psi^{\star}_m\psi_n\right\}:}\ $ is an element
of the group $GL(\infty)$ realized in the infinite-dimensional Grassmannian,
and the normal-ordering should be understood with respect to the vacuum
$|0>$. It was shown in [6] that the solutions to the KP hierarchy are in
one-to-one correspondence with the different group elements $g$.
Thus, we conclude that the group which acts in the space of solutions in the
KP hierarchy case is $GL(\infty)$.

In fact, the same group acts in a little more general case of the
two-dimensional Toda lattice hierarchy. This hierarchy is also satisfied by
the only ($\tau$-)function depending on {\it two} infinite sets of times
$\{t_k\}$, $\{\bar t_k\}$ and one discrete index and is manifestly given by
the ratio
\be
\tau_n(t,\bar t|g)={\left<n|e^{H(t)}ge^{\bar H(\bar
t)}|n\right>\over \left<n|g|n\right>}
\ee
where the second set of times is
coupled to the negative-mode (commuting) Hamiltonians $\bar H(\bar t)=\sum
H_{-k}\bar t_k$ and vacuum states are defined by the conditions
$\psi_m|k>=0,\ \ m<k;\ \ \ \psi^{\star}_m|k>=0,\ \ m\ge k$.
We observe that the space of solutions to the Toda hierarchy is still given
by all the group elements of $GL(\infty)$. However, now we have more flows.
Say, the first non-trivial equation is
\be\label{Toda}
\partial_1\tau_n\bar\partial_1\tau_n-\tau_n\partial_1\bar\partial_1\tau_n=
\tau_{n+1}\tau_{n-1}
\ee

The next group which also has much to do with the space of solutions
often arises as the group describing the reduction of the KP/Toda hierarchy.
Two simplest examples are

1) $\widehat {SL(2)}$-reduction of KP, which corresponds to the KdV hierarchy
(this hierarchy is described by the $\tau$-function depending only on odd
times, which means that, of all infinite set of independent functions $u_i$
of the KP hierarchy, the only function $u_2$ is independent). The same
reduction of the Toda hierarchy gives either Sine-Gordon or Toda chain
hierarchies.  Say, in the Toda chain case, this reduction means that the
$\tau$-function depends only on the difference of times $t_k-\bar t_k$. Then,
making the substitute $e^{\phi_n}\equiv{\tau_{n+1}\over \tau_n}$ in
(\ref{Toda}), one obtains the well-known Toda chain equation
\be
\partial_1^2\phi_n=e^{\phi_{n+1}-\phi_n}-e^{\phi_n-\phi_{n-1}}
\ee
2) $SL(2)$-reduction of the Toda hierarchy. This reduction implies
$\tau_{-n}=0$, $\tau_0=\tau_2=1$, $\tau_{n>2}=0$. Then, one gets
$\partial_1\tau_1\bar\partial_1\tau_1-\tau_1\partial_1\bar\partial_1\tau_1=1$
and, introducing $e^{\phi}\equiv e^{\phi_1-\phi_0}={\f \tau_1^2}$, obtains
the Liouville equation $\partial_1\bar \partial_1\phi=2e^{\phi}$.

Therefore, say, the Liouville system is described by the group pair
$\left(GL(\infty),SL(2)\right)$ (see also [3]).

\bigskip

\noindent
{\bf 3.} Now let us describe the general construction extending the
KP/Toda hierarchies. For doing this, we stress the main two features of the
considered hierarchies which are in charge of all their main properties.
These features are

1) The element $g$ is a group element (of $GL(\infty)$ in our case),
i.e., put it differently, satisfies the co-product relation
$\Delta(g)=g\otimes g$ if one looks at $GL(\infty)$ as Hopf algebra.

2) The fermions are transformed under the group action as
\be\label{BI}
g\psi_ig^{-1}=R_{ik}\psi_k,\ \ \
g\psi^{\star}_ig^{-1}=\psi^{\star}_kR^{-1}_{ki}
\ee
where $R_{ik}$ is a numerical matrix (this can be immediately checked from
the definition of $g$). This means that the fermions are intertwining
operators which intertwine the fundamental representations of $GL(\infty)$.
In fact, if one defines $F_n$ to be the $n$-th fundamental representation,
\be
\psi:\ F_1\otimes F_n\to F_{n+1},\ \ \ \ \psi^{\star}:\ F_{n+1}\to F_n\otimes
F_1
\ee
and index $i$ of fermions runs over the first fundamental representation
$F_1$.

Relation (\ref{BI}) implies that the combination $\sum_i\psi_i\otimes
\psi^{\star}_i$
commutes with $g$. In turn, this results into the infinite set of (bilinear)
identities for the $\tau$-function forming the whole hierarchy of
(KP/Toda) equation [6].

After this preparation, we are ready to consider the general case. Indeed,
let be given a highest-weight representation $\lambda$ of some Lie algebra
${\cal G}$ (indeed, we need the universal enveloping algebra
$U({\cal G})$ and {\it its} representations).
Then, we define $\tau$-function as the generating function of
all matrix elements in the representation $\lambda$:
\be\label{tau}
\tau^{(\lambda)}(t,\bar t|g)\equiv <0|\prod_k e^{t_kT_-^{(k)}}\: g\:
\prod_i e^{\bar t_iT_+^{(i)}}|0>_{\lambda}=\sum <\bn|g|\bm>_{\lambda}
\prod_{i,j}{t_i^{n_i}\bar t_j^{m_j}\over n_i!m_j!}
\ee
where the vacuum state means the highest weight vector, $T_{\pm}^{(k)}$ are
the generators of the corresponding Borel subalgebras of ${\cal G}$ and the
exponentials are supposed to be somehow normal-ordered ($\bn$ ($\bm$) is
vector with the components $n_i$ ($m_i$)).

This general $\tau$-function (\ref{tau}) still satisfies some homogeneous
bilinear identities. To demonstrate this [2], one needs to consider a triple
of representations $V,\hat V,W$ which are intertwined by the operators
\be\label{int}
\Phi:\ \hat V\otimes W\to V,\ \ \ \ \Phi^{\star}:\ V\to
W\otimes \hat V
\ee
Again there is a canonical element which commutes with the group element $g$.
It is provided by the defining property of intertwiner which generalizes
(\ref{BI}):
\be
\Delta (g)\Phi=\Phi g,\ \ \ \Phi^{\star}\Delta (g)=g\Phi^{\star}
\ee

\bigskip

\noindent
{\bf 4.} Now let us say some words of quantizing this system [2,3]. One can
replace the group by the corresponding quantum group and repeat the procedure
of the previous section, but there are some new features to be pointed out:

1) $\tau$-function is no longer commutative. It is still to be defined by
formula (\ref{tau}), with exponential being replaced by the corresponding
$q$-exponentials $e_q$.

2) One needs to differ between left and right intertwiners depending on the
order of spaces in (\ref{int}).

3) The counterpart of the group element $g$ defined by the property
$\Delta (g)=g\otimes g$ {\it does not} belong to the universal enveloping
algebra $U_q({\cal G})$ but to the product $U_q({\cal G})\otimes
U_q^{\star}({\cal G})$, where $U_q^{\star}({\cal G})\equiv A_q(G)$ is the
dual (or, equivalently, the algebra of functions on the quantum group) [3].

To give some example of this last point, let us consider the group
element in more explicit terms [7,3]. In the classical case, it can be
manifestly parametrized as
\be
g=\prod_{ijk}e^{x_-^{(i)}T_-^{(i)}}e^{x_0^{(j)}T_0^{(j)}}e^{x_+^{(k)}T_+^{(k)}}
\ee
where $x^{(i)}$ are the
coordinates on the group manifold. In the quantum group case, these
coordinates become non-commutative and generates the algebra of functions
$A_q(G)$. (In the simplest $A_q(SL(2))$ case, their commutation relations are
$[x_{\pm},x_0]=\left(\log q\right) x_{\pm},\ [x_+,x_-]=0$.)
While in the classical case all the representations
of this algebra are trivial and are given just by the numerical values of the
coordinates $x^{(i)}$, the representations of the quantum algebra are
considerably more involved and, in particular, sometimes
infinite-dimensional.

In the quantum case, the corresponding group element takes the form
\be
g=\prod_{ijk}e_q^{x_-^{(i)}T_-^{(i)}}e^{x_0^{(j)}T_0^{(j)}}
e_{q^{-1}}^{x_+^{(k)}T_+^{(k)}}
\ee
Since the
$\tau$-function (\ref{tau}) is the average of an element from $U_q({\cal
G})\otimes A_q(G)$ over some representation of $U_q({\cal G})$, it is an
element of
the algebra of functions $A_q(G)$, and, therefore, is non-commutative (see 1)
above).

How we already discussed, in the classical case any concrete element $g$
corresponds to some solution to the corresponding classical hierarchy. On the
other hand, the element $g$ is given by some (concrete) values of $x^{(i)}$,
i.e. by some fixed (trivial) representations of the algebra of functions. This
picture is completely extended to the quantum case: {\bf representation of
$A_q(G)$ corresponds to some solution to (quantum) hierarchy}. Certainly, any
reduction, which selects out a subspace in the space of solutions, just
restricts available representations and usually can be described by some
additional group structure.

To conclude this section, let us note that, for the above introduced quantum
group element $g$, there exists the natural group multiplication giving the
(semi)-group structure [3]. Therefore, for the quantized hierarchy, the
underlying (quantum) group still can be described as acting on
the elements $g$, i.e. in the space of solutions to the (quantum) hierarchy.

\bigskip

{\bf 5.}  Now let us consider some examples of the space-time group
structure. For the lack of space, we consider only the quantum case (see
[4] for the classical one and for the proper references). Let us start with
the Liouville quantum mechanics and find the wave function of the Liouville
Hamiltonian.  For doing this, we consider the group element of the {\it
classical} $SL(2)$ group $g=e^{\psi T_-}e^{\phi T_0}e^{\chi T_+}$ and
fix some irreducible spin $j$ representation from the principal (spherical)
series.  We want it to be unitary, i.e. $j+{1\over 2}$ is pure imaginary. Now
take the vectors from this representation satisfying the following properties
\be\label{vects}
<\psi_L|T_-=i\mu_L<\psi_L|,\ \ \ T_+|\psi_R>=i\mu_R|\psi_R>
\ee
The second Casimir operator $C_2=2T_-T_++T_0+{1\over 2}T_0^2=2j(j+1)$ in this
representation. Now we consider the function
$F(\phi)\equiv <\psi_L|e^{\phi T_0}|\psi_R>=e^{-i(\mu_L\psi+\mu_R\chi)}
<\psi_L|g|\psi_R>$. It is called Whittaker function (see references in [4]).
This function is proportional to the Liouville wave function:
\be
2j(j+1)F(\phi)=<\psi_L|e^{\phi T_0}C_2|\psi_R>=
<\psi_L|e^{\phi T_0}\left(2T_-T_++T_0+{1\over 2}T_0^2\right)|\psi_R>=\\=
\left(-2\mu_L\mu_Re^{-2\phi}+{\partial\over\partial\phi}+{1\over 2}
{\partial^2\over\partial\phi^2}\right)F(\phi)
\ee
i.e. $\Psi(\phi)\equiv e^{\phi}F(\phi)$ satisfies
\be
\left({1\over 2}{\partial^2\over\partial\phi^2}-2\mu_R\mu_L\right)\Psi(\phi)=
2\left(j+{1\over 2}\right)^2\Psi(\phi)
\ee
Now to calculate this function, one can take the manifest realization of
$SL(2)$ by the differential operators
\be
T_+ = \frac{\partial}{\partial x},\ T_0=-2x{\d x}
+2j,\ T_-=-x^2{\d x}+2jx
\ee
and, solving differential equations (\ref{vects}), obtain
\be
\psi_R(x)=e^{i\mu_Rx},\ \ \psi_L(x)=x^{-2(j+1)}e^{-{i\mu_L\over x}},
\\ \hbox{i.e.}\ \ \Psi(\phi)=e^{\phi}\int dx x^{-2(j+1)}e^{-{i\mu_L\over x}}
e^{\phi\left(2j-2x{\partial\over\partial x}\right)}e^{i\mu_Rx}\sim
K_{2j+1}\left(2\sqrt{\mu_L\mu_R}e^{-\phi}\right)
\ee
Using this integral representation, one can easily obtain the asymptotics of
the function $\Psi(\phi)$:
\be
\Psi(\phi)\stackreb{\phi\to\infty}{\sim}{1\over\Gamma(1+\nu)}e^{\nu\phi}
-{1\over\Gamma(1-\nu)}e^{-\nu\phi},\ \ \ \nu\equiv 2j+1
\ee
The functions $c_{\pm}(\nu)={1\over\Gamma(1\pm\nu)}$ giving these asymptotics
are called Harish-Chandra functions. Their ratio gives the $S$-matrix and,
equivalently, 2-point function of the theory $S={c_+\over c_-}={\Gamma(1-\nu)
\over\Gamma(1+\nu)}$. \footnote{We omit everywhere from the $S$-matrix
a trivial factor depending on the cosmological constant $\mu$.}

The same procedure can be done for any finite-dimensional group giving [4]
\be
c(\blambda)=\prod_{\balpha\in\Delta^+}{\f \Gamma(1+\bnu\cdot\blambda)}
\ee
where the product runs over all the positive roots of the algebra.
Moreover, it can be continued to the affine case which describes the
Liouville $2d$ quantum field theory. The result is
\be\label{HCHA}
c(\blambda) = \prod_{n\geq 0}
\Gamma^{-1}(p + n\tau)
\prod_{n\geq 1}\Gamma^{-1}(n\tau)\Gamma^{-1}(1-p + n\tau)
\ee
This expression still requires a careful regularization,
but all the infinite products cancel from the corresponding reflection
$S$-matrix
(2-point function)
\be
S(p) = \frac{c(-p)}{c(p)} =
\frac{\Gamma(1+p)\Gamma(1+{p\over\tau})}{\Gamma(1-p)\Gamma(1-{p\over\tau})}
\ee
This result is to be compared with the formulas for the 2-point
functions obtained in papers [8] in a very different
way\footnote{In their notations, $p=2iP/b$ and $\tau=b^2$.}.

To conclude, let us stress again that the space-time group structure
discussed in this section,
although being {\bf classical} one, describes the wave functions
of the {\bf quantum} systems. The same groups also describe the
corresponding classical systems, within a slightly different treatment (see
[4] and references therein).

\bigskip

I would like to thank my colleagues from the mathematical physics group in
ITEP for the valuable discussions and collaboration.

\bigskip

\centerline{{\Large References}}

\begin{enumerate}
\item M.A.Olshanetsky and A.M.Perelomov, {\sl Phys.Rept.,} {\bf 71} (1981)
313

\item A.Gerasimov et al., {\sl Int. J. Mod. Phys.,} {\bf A10} (1995) 2589;
{\bf hep-th 9405011}\\
  S.Kharchev, A.Mironov and A.Morozov, {\bf q-alg/9501013}

\item  A.Mironov, {\bf hep-th/94098190}

\item A.Gerasimov et al., {\bf hep-th/9601161}

\item R.Hirota, in: {\sl Solitons}, R.K.Bullough, P.J.Caudrey, eds.,
Springer-Verlag, NY, 1980

\item  E.Date, M.Jimbo, T.Miwa and M. Kashiwara, in: {\sl Nonlinear Integrable
 Systems}, World Scientific, Singapore, 1983

\item C.Fronsdal and A.Galindo, {\sl Lett.Math.Phys.,} {\bf 27} (1993) 59

\item  H.Dorn, and H.-J.Otto, {\sl Nucl.Phys.,} {\bf B429} (1994) 375;
A.Zamolodchikov \& Al.Zamolodchikov, {\bf hep-th/9506136}

\end{enumerate}
\end{document}